\documentclass[preprint,review,12pt]{elsarticle}

\usepackage{amssymb}

\biboptions{sort&compress,numbers}

\journal{NIMB Nuclear Inst. and Methods in Physics Research, B}

\usepackage{color}
\usepackage{colortbl}

\newcommand{\TV}[1]{}

\begin{document}
\newcommand{\text}[1]{\ensuremath{{\textrm{#1}}}}
\newcommand{\superscript}[1]{\ensuremath{^{\textrm{#1}}}}
\newcommand{\subscript}[1]{\ensuremath{_{\textrm{#1}}}}

\begin{frontmatter}

\title{Modelling surface restructuring by slow highly charged ions}

\author{G.~Wachter$^1$}
\ead{georg.wachter@tuwien.ac.at}
\author{K.~T\H{o}k\'esi$^2$}
\author{G.~Betz$^3$}
\author{C.~Lemell$^1$}
\author{J.~Burgd\"orfer$^1$}

\address{$^1$ Institute for Theoretical Physics, Vienna University of Technology, Wiedner Hauptstra\ss e 8-10, A-1040 Vienna, Austria, EU}
\address{$^2$ Institute of Nuclear Research of the Hungarian Academy of Science (ATOMKI), H-4001 Debrecen, P.O. Box 51, Hungary, EU}
\address{$^3$ Institute for Applied Physics, Vienna University of Technology, Wiedner Hauptstra\ss e 8-10, A-1040 Vienna, Austria, EU}

\begin{abstract}
We theoretically investigate surface modifications on alkaline earth halides due to highly charged ion impact, focusing on recent experimental evidence for both etch pit and nano-hillock formation on CaF$_{\mathrm{2}}$ \cite{Said2012Phase}. We discuss mechanisms for converting the  projectile potential and kinetic energies into thermal energy capable of changing the surface structure. A proof-of-principle classical molecular dynamics simulation suggests the existence of two thresholds which we associate with etch pit and nano-hillock formation in qualitative agreement with experiment. 

\end{abstract}

\begin{keyword}
surface modification \sep nano hillock \sep etch pit \sep molecular dynamics 
\end{keyword}

\end{frontmatter}



\section{Introduction}  
\label{sec:intro}
Investigations of the interactions of slow ($v < v_{\mathrm{Bohr}}$), highly charged ions (HCI) with surfaces have aimed at gaining an improved understanding of the dynamics  governing the neutralization and eventual dissipation of the large projectile potential energy in a nano scale region of the target on an ultrashort time scale of a few femtoseconds (see \cite{Gillaspy2001Highly,Aumayr2008Nanosized} and references therein). The projectile potential energy is given by the sum over the sequential ionization energies up to the charge state $q$ and can easily reach 5--50 keV for projectile charge states $q \sim $ 20 -- 50. The resulting nano-sized surface modifications may lead to applications in material science \cite{Pomeroy2007Selectable}. Different topographies of surface modifications have been observed, depending on the projectile potential and kinetic energy and the target material \cite{Said2008Creation,Tona2008Electronic} (see \cite{Aumayr2011Single} for a recent review).

Even for seemingly similar wide-bandgap insulator targets, namely ionic crystals of alkaline earth halides and alkali halides, experiments have found different types of surface modifications. On KBr, pits of one atomic layer depth have been observed \cite{Heller2008Defect} while on CaF$_{\mathrm{2}}$ nano-meter sized hillocks protruding from the surface have been found \cite{Said2008Creation}. In a recent contribution \cite{Said2012Phase}, pits have also been found on CaF$_{\mathrm{2}}$ surfaces if the sample was etched after irradiation (``etch pits''). The appearance of hillocks and etch pits as a function of both potential and kinetic energies of the impinging projectiles suggests the existence of three regions in parameter space: for low kinetic and potential energies, no surface modifications are observed while for a combination of larger kinetic and potential energies, only etch pits but no hillocks are  found. Finally, for sufficiently large projectile potential energy both nano-hillocks and etch pits are observed only weakly dependent on the concomitant kinetic energy. The appearance of these two different surface modifications for the same target material was associated with the existence of two different thresholds, one for etchable damage and one for hillock formation.

The involved complex many-body dynamics spanning many orders of magnitude in time pose a considerable challenge to theory. In this contribution, we present a first attempt of a multi-scale modelling making use of the different times scales rendering a qualitative, yet microscopic, model of surface modifications.

The plan for this paper is as follows. We briefly review theoretical models for the transfer of projectile potential and kinetic energy to the electronic and subsequently to the lattice degree of freedom. Then, we present a proof-of-principle molecular dynamics simulation with the heated nanoscale surface region as initial condition that qualitatively reproduces the existence of two thresholds associated with the creation of etchable defects and surface restructuring in the form of nano-hillocks in agreement to experiment. Atomic units are used throughout unless stated otherwise.

\section{Scenario for energy deposition} 
\label{sec:review}
The multi-scale model for the energy deposition by slow highly charged ions on a CaF$\mathrm{_2}$ target is based on the scenario in which the initial deposition into the electronic degree of freedom and subsequently into the lattice degree of freedom proceeds on shorter time scales than the subsequent atomic rearrangements and structural changes \cite{Kaganov1957Relaxation,Toulemonde1992Transient,Lemell2007Nanohillock,Said2008Creation}. This allows to treat such processes separately. 

As the slow highly charged ion approaches the surface, the primary energy conversion from potential energy of the projectile to electronic energy of the target proceeds by a complex de-excitation cascade, both above and below the surface  \cite{Burgdorfer1991Abovesurface,Burgdorfer1995Fast,Hagg1997Abovesurface,Wirtz2003Liouville,Lemell2007Nanohillock}. As the ion approaches the surface, electron capture to highly excited projectile states sets in, followed by a subsequent de-excitation by Auger decay processes and radiative decay. Close to the surface, the self-image interaction leads to an upwards shift in energy of the highly excited states, leading to re-ionization of the projectile. As a consequence, a large fraction of projectiles are still highly charged upon entering the first few target layers. De-excitation below the surface continues with states below or in resonance with the target valence states, leading to moderately excited projectile states. They decay by Auger sequences emitting electrons with low to intermediate energies up to a few hundred eV. In the case of inner-shell holes (e.g., Ar$^{\mathrm{17+}}$), a small number of keV electrons are released. The kinetic energy of the projectile determines the depth at which the neutralization sequence is complete, with penetration depths of a few nm for kinetic energies of $\lesssim 10^{5}$ eV \cite{Said2008Creation}. The process taking place within the first $\sim$ 20 femtoseconds of the interaction can be simulated using the classical-over-the-barrier model \cite{Burgdorfer1991Abovesurface} and its extension to insulator surfaces \cite{Burgdorfer1995Fast}. A quantitative analysis of slow highly charged ion impact on CaF$_{\mathrm{2}}$ is given in \cite{Lemell2007Nanohillock}, relating the charge state of the projectile to the electron emission spectrum around the impact area. 

\begin{figure}[!ht] 
  \centering
  \includegraphics[width=\columnwidth,natwidth=1476bp,natheight=473bp] {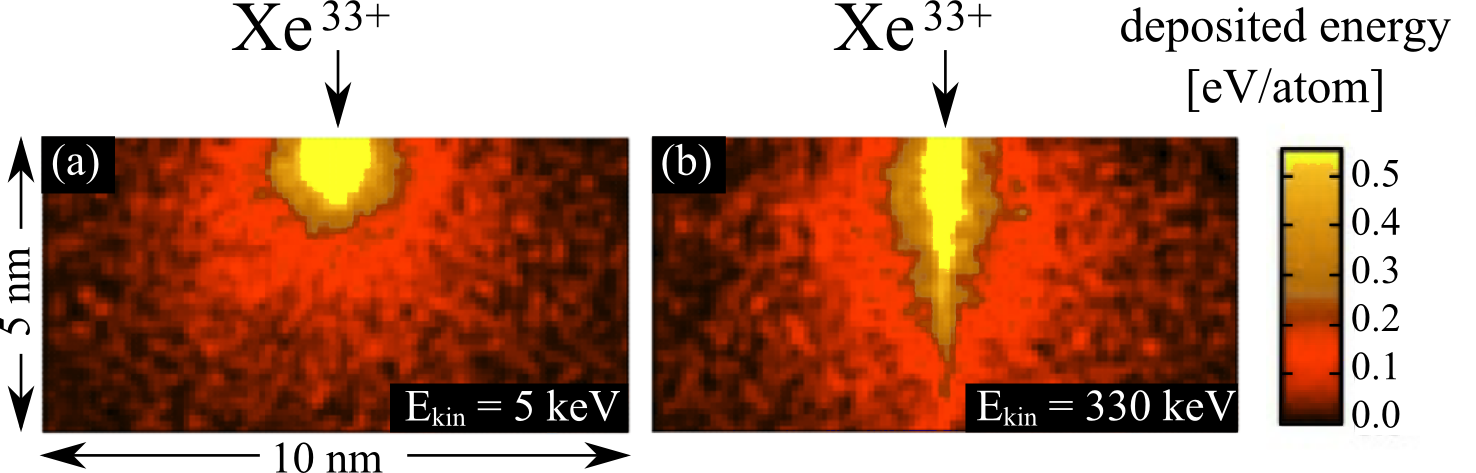} 
  \caption{ \label{fig:flames}
Simulated thermal energy deposition due to impact of Xe$^{33+}$ for low ($E_{\mathrm{kin}}=5$ keV, left) and larger ($E_{\mathrm{kin}}=330$ keV, right) projectile kinetic energy. For details see text and \cite{Said2008Creation}. 
  }
\end{figure}

These liberated electrons lead to the subsequent electron cascade, transferring part of their kinetic energy to the target lattice by electron-phonon coupling. The relevant time scale for this process is a few hundred femtoseconds. This electron cascade can be analyzed by a classical trajectory Monte Carlo simulation taking into account elastic and inelastic scattering events, leading to the creation of secondary electrons and to excitations of phonons in the interaction with crystal atoms \cite{Lemell2007Nanohillock,Said2008Creation}. One key result, somewhat counterintuitive at first sight, is that the low-energy electrons ($\lesssim $ 200 eV) are far more effective in the local heating of the target lattice than the energetic electrons ($\sim$ keV). Even though the available energy of the low-energy electrons is lower, their contribution to the heating of the lattice near the impact area is larger because they deposit their energy in close proximity to the impact site leading to a large energy \emph{ density } whereas high-energy electrons transport and spread their energy over a much larger volume due to their larger mean free path. One consequence of this rapid diffusion is that only a part of the projectile energy is effectively deposited as heat near the impact site. Typically, for an  initial potential energy of the order of $\sim 10$ keV (corresponding to Xe$^{\mathrm{24+}}$) only 20 \% are released as kinetic energy of low energy electrons, part of which is deposited as heat near the impact site. In total, an energy density near the melting point is reached in a region of diameter $\sim 3$ nm (fig.~\ref{fig:flames}). The influence of the kinetic energy on the potential energy deposition is primarily to control the shape and size of the heated ``core'' volume of the impact site \cite{Said2008Creation}. The shape varies from hemispherical for slow ions with smaller penetration depth (fig.~\ref{fig:flames}a) to a flame-like shape for faster ions with larger penetration depth (fig.~\ref{fig:flames}b). In addition, the energy loss of the ion due to its nuclear stopping power can contribute to heating as well. In the present scenario, this contribution to local heating is of minor importance. At projectile energies of a few hundred keV (right panel in fig.~\ref{fig:flames}), the nuclear stopping power in the bulk is near its maximum of $2.5$ keV/nm. The energy is transferred to the target via a collision cascade diffusing the energy around the projectile path and transporting it away from the surface and into the bulk. SRIM calculations \cite{Ziegler2010SRIM} show that the deposited energy density near the surface is too low to induce surface restructuring on its own but can synergistically work together with the electron cascade to thermally activate the lattice. 

\section{Simulation of target restructuring}
\label{sec:md}

The physical scenario laid out in the preceding section results in strong local heating in a region near the impact point. In the following, we assume that heating and subsequent melting is the main driving force for defect formation. This assumption is corroborated by earlier analyses linking nano-hillock formation following HCI impact to a thermal spike picture \cite{Lemell2007Nanohillock} based on the two-temperature model \cite{Kaganov1957Relaxation}. We parametrize the thermal excitation due to HCI impact by the linear size of the impact region, its radius $R$ (assuming a hemispherical shape), and its effective temperature, $T$. These parameters can account for both electronic and nuclear contributions to local heating. The parameters $R$ and $T$ serve as input for our proof-of-principle molecular dynamics (MD) simulation by which we explore the mechanisms leading to melting, etchability, and eventually surface restructuring. We note that in this work for simplicity we do not include deviations from a hemispherical shape of the hot zone and from a homogeneous temperature distribution within the hemisphere to keep the number of parameters minimal. Test runs for other shapes and smeared out temperature distributions yield approximately the same results.

The target atoms are described as point particles interacting via static ionic pair potentials with short range admixtures of the Born-Mayer-Huggins form,
\begin{equation}
\label{eq:atompot}
V_{ij}(r) = Z_i Z_j e^2 / r + A_{ij} e^{-r / \rho_{ij} } - C_{ij} / r^6
\end{equation}
where we assume the bare ionic charges $Z_{\mathrm{Ca}}=+2$ and $Z_{\mathrm{F}}=-1$ using  parameters from \cite{Gillan1986Collective} which are tuned to give the correct lattice constant and F vacancy formation energy (Frenkel defect energy) at low temperatures. In line with our assumption that the damage mechanism is predominantly thermal \cite{Lemell2007Nanohillock}, we neglect the preceding de-excitation and charge transfer processes and assume that all atoms are in their electronic ground state, giving an upper bound for crystal stability. Similar approaches have been used to analyze  the formation of tracks by swift heavy ion impact, where strong local heating due to electronic excitations also leads to melting and restructuring \cite{Kluth2008Fine,Moreira2010Atomistic}.

The size of the simulation box was up to 13.9$\times$11.4$\times$4.6 nm$^3$ ($\sim 55000$ atoms). Periodic boundary conditions are assumed perpendicular to the surface to avoid edge effects. The evaluation of Coulomb forces is performed in a shifted-force scheme (``Wolf method'' \cite{Wolf1992Reconstruction,Wolf1999Exact,Fennell2006Ewald}) with a cut-off of 0.9 nm, the accuracy of which has been demonstrated for a variety of systems including ionic crystals \cite{Zahn2002Enhancement,Fennell2006Ewald} . We use a Berendsen thermostat \cite{Berendsen1984Molecular} to thermalize an ensemble of atoms. The outermost layers of the simulation box can be coupled to a heat bath at room temperature to mimic the influence of the remaining crystal and act as a heat sink. The simulation reproduces a range of ground state and equilibrium properties of CaF$_\mathrm{2}$ such as superionicity \cite{Lindan1991Molecular}, sound velocity, melting point, and thermal diffusivity \cite{Andersson1987Thermal}, which is a key quantity for the transport of heat away from the impact region. One major limitation of the present simulation is that the atomic potentials (eq.~\ref{eq:atompot}) do not allow for charge transfer and transitions between potential hypersurfaces giving rise to chemical reactions. Therefore, the simulation results are intended to give primarily qualitative insights. 

\begin{figure}[!ht] 
  \centering
  \includegraphics[width=0.8 \columnwidth,natwidth=764bp,natheight=1100bp] {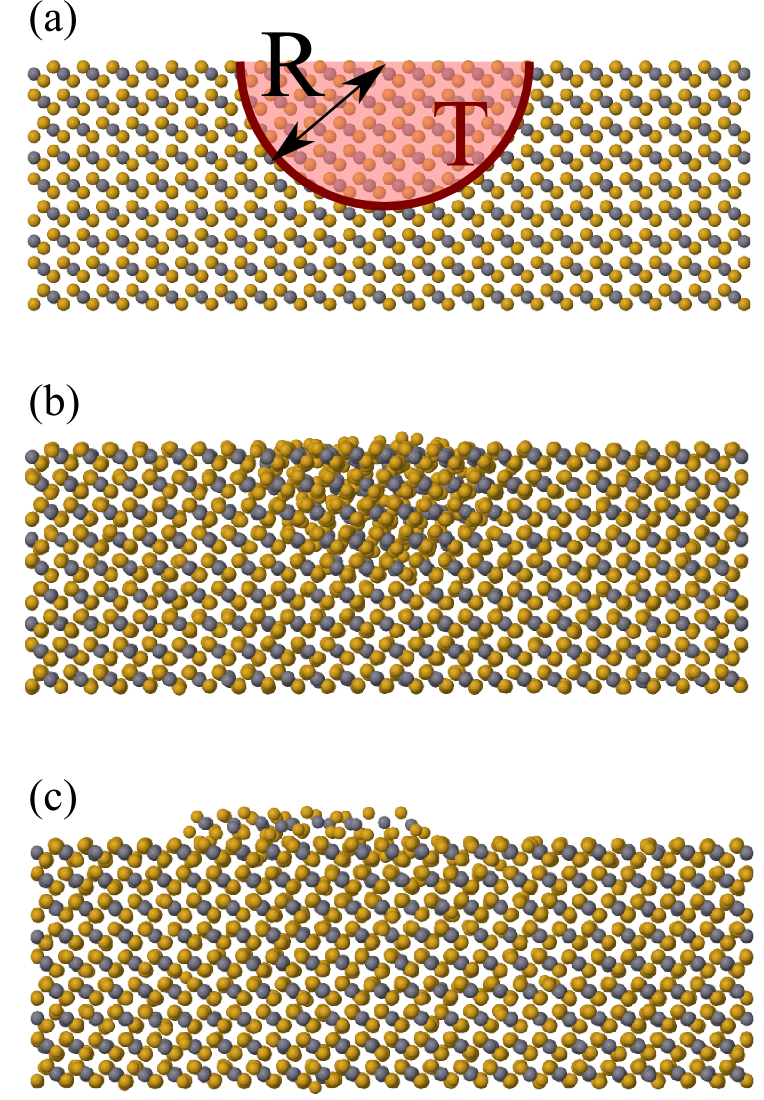} 
  \caption{ \label{fig:mdsim}
Simulation of etchable defects and hillock formation. (a) Shape of the hemispherical region (radius $R$) heated to a temperature $T$ in a CaF$_{\mathrm{2}}$ crystal (grey: Ca, yellow: F). (b) Snapshot after the heating (duration 100 fs). (c) The cooled crystal after 15 ps simulation time shows few dislocations in the bulk and a patchy triple layer on the surface is visible as a hillock under the atomic force microscope.  }
\end{figure}

The virtual crystal is first equilibrated at a temperature of 300 K for 3 ps. To simulate the effect of a highly charged ion impact and the resulting heating of the target near the impact point, a central hemisphere (radius $R$) is heated to a temperature $T$ within 100 fs, the time scale of the  electron cascade (fig.~\ref{fig:mdsim}). A simulation snapshot after the heating is displayed in fig.~\ref{fig:mdsim} (b). We observe a large number of thermally activated F$^{-}$. Within our simulation, most of the activated F ions find their way back to F vacancy sites as the system cools down (simulation time 15 ps, simulation snapshot in fig.~\ref{fig:mdsim} (c)), which is to be expected for a non-amorphizable material such as CaF$_\mathrm{2}$ \cite{Toulemonde2000Transient}. This reconstitution is, however, in part an artifact due to the absence of reactive scattering processes which are difficult to incorporate into a classical molecular dynamics simulation in a consistent way \cite{Chen2007QTPIE}. For example, the thermally activated F$^{-}$ ions can form fluorine gas by the charge transfer reaction Ca$^{2+}$ + 2 F$^{-}$ $\to$ Ca $+$ F$_2$. Similarly, neutral F$^0$ ``holes'' created in the F sub-lattice can covalently bond 2 F$^{0}$ $\to$ F$_2$, even without need for a nearby Ca. The newly formed F$_2$ molecules could aggregate as interstitials and evaporate, leaving behind Ca aggregates near the surface constituting the hillock, similar to bulk restructuring observed in CaF$_{\mathrm{2}}$ after swift heavy ion impact \cite{Chadderton2003Nuclear,Jensen1998Microscopic}. 

Depending on $R$ and $T$, the formation of hillocks can be observed (fig.~\ref{fig:mdsim} (c)) as a patchy partial triple layer on the surface of the  crystal. To quantify the sensitivity to etching, we determine the number of thermally activated atoms which can be associated with the size of the molten core region. This is done by counting the number of F ions that end up further than half a nearest-neighbor distance of the F-sublattice ($\approx 0.14$ nm) away from their original position, i.e.~atoms which have moved to another lattice site during the course of the simulation are counted as ``activated''. We assume that a large number of thermally activated atoms weakens the target structure and leads to increased sensitivity to etching. Obviously, sensitivity to etching can only be expected if the defects are created at or near the surface. In our simulations, we observe that atoms are  thermally activated preferentially near the surface (compare also fig.~\ref{fig:hillpic}).

\begin{figure*}[!ht] 
  \centering
  \includegraphics[width=\textwidth,natwidth=2433bp,natheight=833bp] {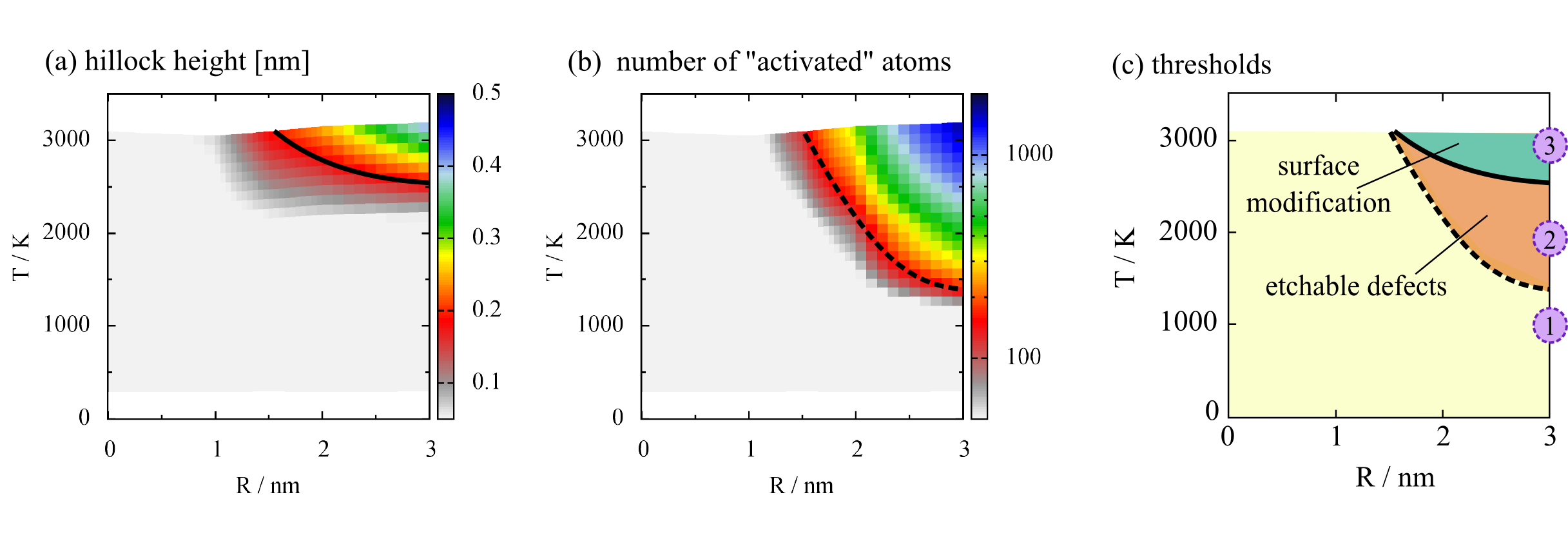} 
  \caption{ \label{fig:thresholds}
Simulation results as a function of MD parameters $R$ and $T$ respectively describing the size and temperature of the thermally activated region created by the preceding HCI impact. (a) Height of simulated hillocks. (b) Number of thermally activated atoms surmised to lead to etchable defects (details see text). (c) Combined graph showing two different thresholds, for example along the $R=3$ nm contour marked by points 1, 2 and 3 corresponding to simulation results shown in fig. \ref{fig:hillpic}.
  }
\end{figure*}

\begin{figure}[!ht] 
  \centering
  \includegraphics[width=\columnwidth,natwidth=1340bp,natheight=1368bp] {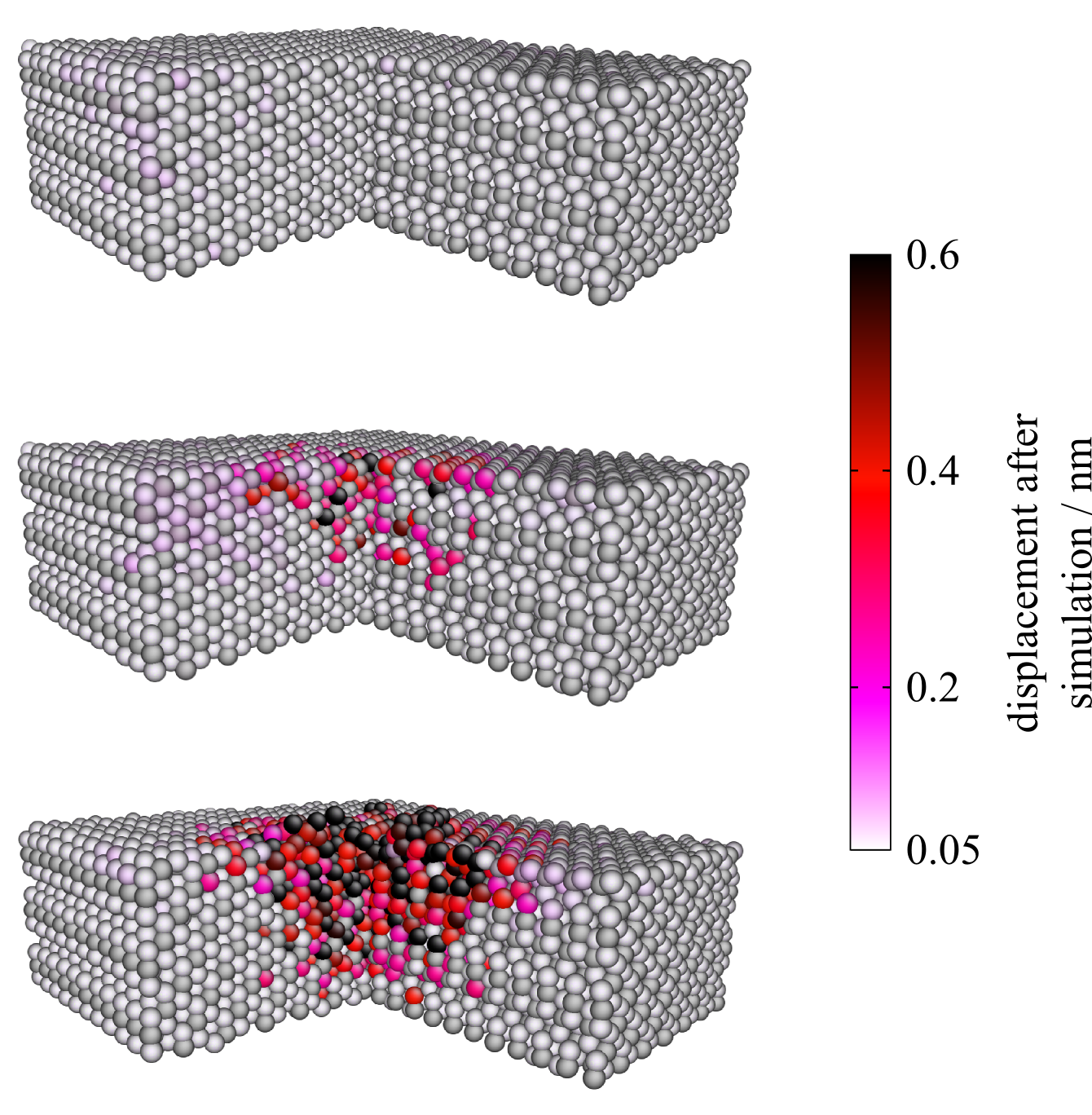} 
  \caption{ \label{fig:hillpic}
Typical simulation results showing the transition from no restructuring (top) to a large number of activated atoms near the impact region (center) to surface modification (bottom). Light atoms are F while darker atoms are Ca. The distance of each atom to its original position at the beginning of the simulation is overlayed as color code. The simulations correspond to $R=3$ nm and $T=$ 1000 K, 2000 K and 3000 K respectively, marked by point labels 1, 2 and 3 in fig. \ref{fig:thresholds}(c). Figure created with AtomEye \cite{Li2003AtomEye}.
  }
\end{figure}

We consider now the dependence of formation of hillocks and of etchable defects as a function of $R$ and $T$ describing the width and temperature of the heated zone. Fig.~\ref{fig:thresholds} shows contour maps of the simulated height of hillocks (panel (a), linear color scale) and the number of thermally activated atoms in the simulation box (panel (b), logarithmic color scale). For both observables thresholds are identified (marked by black lines). Since both hillock height and the number of thermally activated atoms are statistical quantities the results present an average over 10 simulation runs each on a grid of 5$\times$5 points and interpolated in between. The two thresholds for hillock formation on one hand and etchable defects on the other hand are compared in fig.~\ref{fig:thresholds} (c). In the experiment, hillock formation is always accompanied by etchability \cite{Said2012Phase}, consistent with the present simulation where a large number of activated atoms is visible beneath a hillock (compare fig.~\ref{fig:thresholds}a, \ref{fig:thresholds}b and right-hand panel of fig.~\ref{fig:hillpic}). 

Typical snapshots after 15 ps for three different $T$ at fixed $R=$3 nm (marked by points 1,2, and 3 in fig.~\ref{fig:thresholds}) show that no damage is visible for low temperatures (fig.~\ref{fig:hillpic} top panel), while a number of activated atoms, consisting mainly of F atoms, concentrated around the impact point appears for moderate temperatures (fig.~\ref{fig:hillpic} center panel). For high temperatures (fig.~\ref{fig:hillpic} bottom panel) the number of activated atoms is even larger now also including Ca atoms near the impact point. A hillock protruding from the surface is observed with a typical height of one triple layer ( $\sim$0.3 nm). 

The two thresholds observed in the simulation are consistent with the ones observed in the experiment \cite{Said2012Phase}. For a more detailed quantitative connection the relationship between the experimentally controlled projectile parameters (potential and kinetic energies) with the MD simulation parameters $R$ and $T$ would be required. While the dependence of $R$ and $T$ on the potential energy can be approximately determined within our simulation for the initial heating by hot electrons, the dependence on the kinetic energy is more involved as increased energy dispersion (see fig.~\ref{fig:flames}) is partially offset by increased nuclear stopping. The combined effect of both the de-excitation cascade and the collision cascade is currently not included in our heat deposition model. This exact relationship is, however, not of importance for the qualitative observation of two thresholds, which are crossed by any path leading from low values of $(R,T)$ to large $(R,T)$ fig.~\ref{fig:thresholds} (c).

\begin{figure*}[!ht] 
  \centering
  \includegraphics[width=\textwidth,natwidth=2314bp,natheight=889bp] {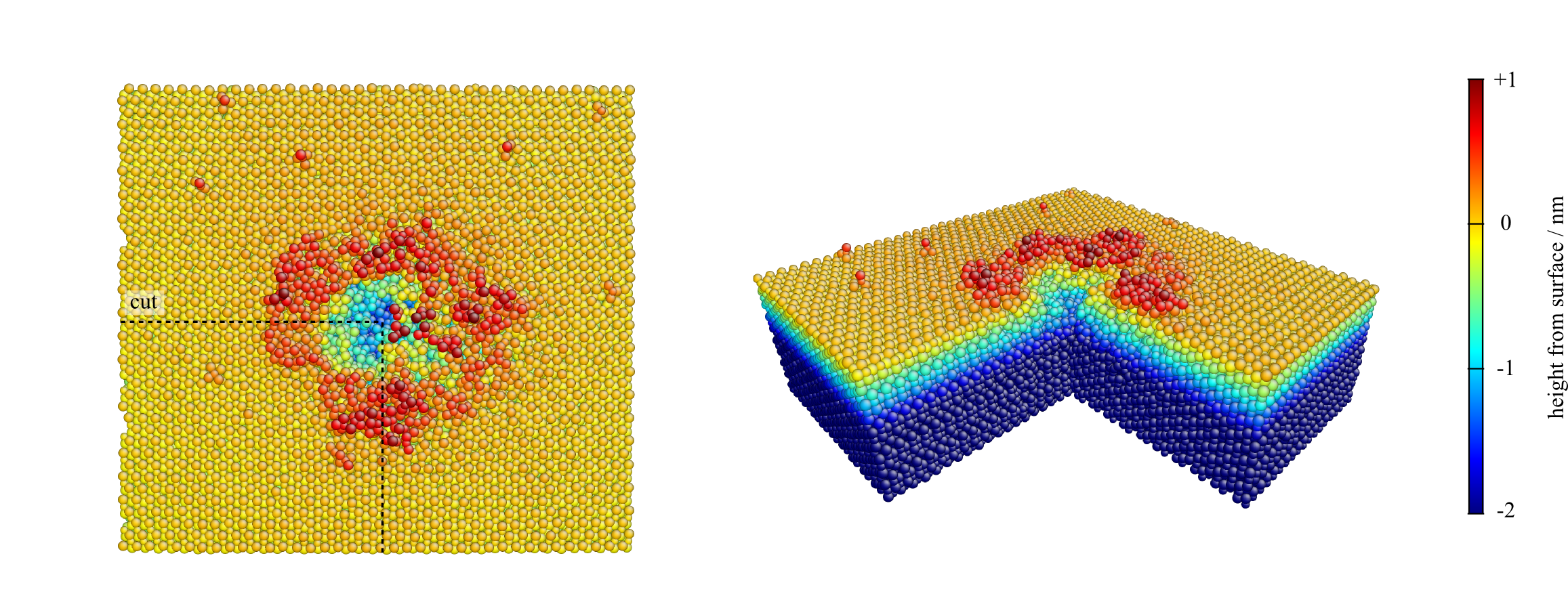} 
  \caption{ \label{fig:crater}
Snapshot of MD simulation displaying crater formation. The maximum temperature of the central hemisphere (radius $R = 2$ nm) was taken as $T = 18000$ K. Left: Top-down view, right: cut through center, with color according to the distance from the surface layer. System size is 14.7$\times$13.4$\times$4.5 nm$^3$ (68400 atoms). Figure created with AtomEye \cite{Li2003AtomEye}. 
  }
\end{figure*}

For future nano-structuring applications it is of interest to explore regions of larger $T$. The latter could, possibly, be reached by more highly charged ions (e.g.~U$^{\mathrm{92+}}$) and/or faster projectiles. With a five-fold increase in temperature (deposited thermal energy) compared with the one from nano-hillock formation discussed above such that the sublimation threshold is reached we observe a pronounced crater, i.e.~a combined hillock and pit (fig.~\ref{fig:crater}). The depth of the crater is $\sim 1.5$ nm corresponding to a sputter yield of $\sim 440$ atoms (volume of the crater). In addition, a swollen region forming the elevated crater rim (height $\sim 1$ nm) is visible. The total thermal energy deposited in the simulation is $\sim 6$ keV, which might be reached with the highest charge states or, more likely, with swift heavy ions.

\section{Conclusions and outlook}
In this contribution, we have theoretically investigated surface restructuring by highly charged ions. We have reviewed the mechanisms underlying the surface restructuring and the theoretical tools to model the deposition of the potential energy of the highly charged ion to the target lattice. Within a proof-of-principle molecular dynamics simulation to estimate the effects of the HCI impact on the target lattice we could identify the appearance of two thresholds associated with etchability and with nano-hillock formation in qualitative agreement with experiment. 

To improve the current theoretical model, the inclusion of charge transfer reaction processes and chemical changes during the de-excitation sequence is of paramount importance. While the formation of surface modification happens on a time scale of a few picoseconds, the chemical structure of the surface and bulk continues to change through defect agglomeration and defect mediated desorption on a longer time scale. 

Along these lines, measurements of angle- and species-resolved sputter yields could provide important clues as to the microscopic dynamics of the nano-melt during hillock formation. An in-situ measurement of the chemical composition of hillocks would give information about the final state of the system as well as a more complete characterization of hillock formation for potential applications in nano-science and could provide important benchmarks for MD simulations that include reactive processes.

\section*{Acknowledgments}
This work was supported by the Austrian Science Foundation FWF under Proj.\ Nos.\ SFB-041 ViCoM and P21141-N16. G.W.\ thanks the International Max Planck Research School of Advanced Photon Science for financial support. The authors would like to thank F.\ Aumayr for fruitful discussions.


\bibliographystyle{model1-num-names}
\bibliography{georgwachterclean}

\begin{thebibliography}{31}
\expandafter\ifx\csname natexlab\endcsname\relax\def\natexlab#1{#1}\fi
\providecommand{\bibinfo}[2]{#2}
\ifx\xfnm\relax \def\xfnm[#1]{\unskip,\space#1}\fi
\bibitem[{Said et~al.(2012)Said, Wilhelm, Heller, Facsko, Lemell, Wachter,
  Burgd\"{o}rfer, Ritter, and Aumayr}]{Said2012Phase}
\bibinfo{author}{A.~S.~E. Said}, \bibinfo{author}{R.~A. Wilhelm},
  \bibinfo{author}{R.~Heller}, \bibinfo{author}{S.~Facsko},
  \bibinfo{author}{C.~Lemell}, \bibinfo{author}{G.~Wachter},
  \bibinfo{author}{J.~Burgd\"{o}rfer}, \bibinfo{author}{R.~Ritter},
  \bibinfo{author}{F.~Aumayr},
\newblock \bibinfo{journal}{Phys. Rev. Lett.} \bibinfo{volume}{109}
  (\bibinfo{year}{2012}) \bibinfo{pages}{117602}.
\bibitem[{Gillaspy(2001)}]{Gillaspy2001Highly}
\bibinfo{author}{J.~D. Gillaspy},
\newblock \bibinfo{journal}{J. Phys. B: At. Mol. Opt. Phys.}
  \bibinfo{volume}{34} (\bibinfo{year}{2001}) \bibinfo{pages}{R93--R130}.
\bibitem[{Aumayr et~al.(2008)Aumayr, Elsaid, and Meissl}]{Aumayr2008Nanosized}
\bibinfo{author}{F.~Aumayr}, \bibinfo{author}{A.~Elsaid},
  \bibinfo{author}{W.~Meissl},
\newblock \bibinfo{journal}{Nucl. Instrum. Methods Phys. Res., Sect. B}
  \bibinfo{volume}{266} (\bibinfo{year}{2008}) \bibinfo{pages}{2729--2735}.
\bibitem[{Pomeroy et~al.(2007)Pomeroy, Grube, Perrella, and
  Gillaspy}]{Pomeroy2007Selectable}
\bibinfo{author}{J.~M. Pomeroy}, \bibinfo{author}{H.~Grube},
  \bibinfo{author}{A.~C. Perrella}, \bibinfo{author}{J.~D. Gillaspy},
\newblock \bibinfo{journal}{Applied Physics Letters} \bibinfo{volume}{91}
  (\bibinfo{year}{2007}) \bibinfo{pages}{073506+}.
\bibitem[{Said et~al.(2008)Said, Heller, Meissl, Ritter, Facsko, Lemell,
  Solleder, Gebeshuber, Betz, Toulemonde, M\"{o}ller, Burgd\"{o}rfer, and
  Aumayr}]{Said2008Creation}
\bibinfo{author}{A.~S.~E. Said}, \bibinfo{author}{R.~Heller},
  \bibinfo{author}{W.~Meissl}, \bibinfo{author}{R.~Ritter},
  \bibinfo{author}{S.~Facsko}, \bibinfo{author}{C.~Lemell},
  \bibinfo{author}{B.~Solleder}, \bibinfo{author}{I.~C. Gebeshuber},
  \bibinfo{author}{G.~Betz}, \bibinfo{author}{M.~Toulemonde},
  \bibinfo{author}{W.~M\"{o}ller}, \bibinfo{author}{J.~Burgd\"{o}rfer},
  \bibinfo{author}{F.~Aumayr},
\newblock \bibinfo{journal}{Phys. Rev. Lett.} \bibinfo{volume}{100}
  (\bibinfo{year}{2008}) \bibinfo{pages}{237601}.
\bibitem[{Tona et~al.(2008)Tona, Fujita, Yamada, and
  Ohtani}]{Tona2008Electronic}
\bibinfo{author}{M.~Tona}, \bibinfo{author}{Y.~Fujita},
  \bibinfo{author}{C.~Yamada}, \bibinfo{author}{S.~Ohtani},
\newblock \bibinfo{journal}{Phys. Rev. B: Condens. Matter} \bibinfo{volume}{77}
  (\bibinfo{year}{2008}) \bibinfo{pages}{155427}.
\bibitem[{Aumayr et~al.(2011)Aumayr, Facsko, El-Said, Trautmann, and
  Schleberger}]{Aumayr2011Single}
\bibinfo{author}{F.~Aumayr}, \bibinfo{author}{S.~Facsko},
  \bibinfo{author}{A.~S. El-Said}, \bibinfo{author}{C.~Trautmann},
  \bibinfo{author}{M.~Schleberger},
\newblock \bibinfo{journal}{J. Phys.: Condens. Matter} \bibinfo{volume}{23}
  (\bibinfo{year}{2011}) \bibinfo{pages}{393001}.
\bibitem[{Heller et~al.(2008)Heller, Facsko, Wilhelm, and
  M\"{o}ller}]{Heller2008Defect}
\bibinfo{author}{R.~Heller}, \bibinfo{author}{S.~Facsko},
  \bibinfo{author}{R.~A. Wilhelm}, \bibinfo{author}{W.~M\"{o}ller},
\newblock \bibinfo{journal}{Phys. Rev. Lett.} \bibinfo{volume}{101}
  (\bibinfo{year}{2008}) \bibinfo{pages}{096102}.
\bibitem[{Kaganov et~al.(1957)Kaganov, Lifshitz, and
  Tanatarov}]{Kaganov1957Relaxation}
\bibinfo{author}{M.~I. Kaganov}, \bibinfo{author}{I.~M. Lifshitz},
  \bibinfo{author}{L.~V. Tanatarov},
\newblock \bibinfo{journal}{Sov. Phys. JETP} \bibinfo{volume}{4}
  (\bibinfo{year}{1957}) \bibinfo{pages}{173}.
\bibitem[{Toulemonde et~al.(1992)Toulemonde, Dufour, and
  Paumier}]{Toulemonde1992Transient}
\bibinfo{author}{M.~Toulemonde}, \bibinfo{author}{C.~Dufour},
  \bibinfo{author}{E.~Paumier},
\newblock \bibinfo{journal}{Phys. Rev. B: Condens. Matter} \bibinfo{volume}{46}
  (\bibinfo{year}{1992}) \bibinfo{pages}{14362}.
\bibitem[{Lemell et~al.(2007)Lemell, El-Said, Meissl, Gebeshuber, Trautmann,
  Toulemonde, Burgd\"{o}rfer, and Aumayr}]{Lemell2007Nanohillock}
\bibinfo{author}{C.~Lemell}, \bibinfo{author}{A.~S. El-Said},
  \bibinfo{author}{W.~Meissl}, \bibinfo{author}{I.~C. Gebeshuber},
  \bibinfo{author}{C.~Trautmann}, \bibinfo{author}{M.~Toulemonde},
  \bibinfo{author}{J.~Burgd\"{o}rfer}, \bibinfo{author}{F.~Aumayr},
\newblock \bibinfo{journal}{Solid-State Electron.} \bibinfo{volume}{51}
  (\bibinfo{year}{2007}) \bibinfo{pages}{1398--1404}.
\bibitem[{Burgd\"{o}rfer et~al.(1991)Burgd\"{o}rfer, Lerner, and
  Meyer}]{Burgdorfer1991Abovesurface}
\bibinfo{author}{J.~Burgd\"{o}rfer}, \bibinfo{author}{P.~Lerner},
  \bibinfo{author}{F.~W. Meyer},
\newblock \bibinfo{journal}{Phys. Rev. A: At. Mol. Opt. Phys.}
  \bibinfo{volume}{44} (\bibinfo{year}{1991}) \bibinfo{pages}{5674--5685}.
\bibitem[{Burgd\"{o}rfer et~al.(1995)Burgd\"{o}rfer, Reinhold, and
  Meyer}]{Burgdorfer1995Fast}
\bibinfo{author}{J.~Burgd\"{o}rfer}, \bibinfo{author}{C.~Reinhold},
  \bibinfo{author}{F.~Meyer},
\newblock \bibinfo{journal}{Nucl. Instrum. Methods Phys. Res., Sect. B}
  \bibinfo{volume}{98} (\bibinfo{year}{1995}) \bibinfo{pages}{415--419}.
\bibitem[{H\"{a}gg et~al.(1997)H\"{a}gg, Reinhold, and
  Burgd\"{o}rfer}]{Hagg1997Abovesurface}
\bibinfo{author}{L.~H\"{a}gg}, \bibinfo{author}{C.~O. Reinhold},
  \bibinfo{author}{J.~Burgd\"{o}rfer},
\newblock \bibinfo{journal}{Phys. Rev. A: At. Mol. Opt. Phys.}
  \bibinfo{volume}{55} (\bibinfo{year}{1997}) \bibinfo{pages}{2097--2108}.
\bibitem[{Wirtz et~al.(2003)Wirtz, Reinhold, Lemell, and
  Burgd\"{o}rfer}]{Wirtz2003Liouville}
\bibinfo{author}{L.~Wirtz}, \bibinfo{author}{C.~O. Reinhold},
  \bibinfo{author}{C.~Lemell}, \bibinfo{author}{J.~Burgd\"{o}rfer},
\newblock \bibinfo{journal}{Phys. Rev. A: At. Mol. Opt. Phys.}
  \bibinfo{volume}{67} (\bibinfo{year}{2003}) \bibinfo{pages}{012903}.
\bibitem[{Ziegler et~al.(2010)Ziegler, Ziegler, and Biersack}]{Ziegler2010SRIM}
\bibinfo{author}{J.~F. Ziegler}, \bibinfo{author}{M.~D. Ziegler},
  \bibinfo{author}{J.~P. Biersack},
\newblock \bibinfo{journal}{Nucl. Instrum. Methods Phys. Res., Sect. B}
  \bibinfo{volume}{268} (\bibinfo{year}{2010}) \bibinfo{pages}{1818--1823}.
\bibitem[{Gillan(1986)}]{Gillan1986Collective}
\bibinfo{author}{M.~J. Gillan},
\newblock \bibinfo{journal}{J. Phys. C: Solid State Phys.} \bibinfo{volume}{19}
  (\bibinfo{year}{1986}) \bibinfo{pages}{3391}.
\bibitem[{Kluth et~al.(2008)Kluth, Schnohr, Pakarinen, Djurabekova, Sprouster,
  Giulian, Ridgway, Byrne, Trautmann, Cookson, Nordlund, and
  Toulemonde}]{Kluth2008Fine}
\bibinfo{author}{P.~Kluth}, \bibinfo{author}{C.~S. Schnohr},
  \bibinfo{author}{O.~H. Pakarinen}, \bibinfo{author}{F.~Djurabekova},
  \bibinfo{author}{D.~J. Sprouster}, \bibinfo{author}{R.~Giulian},
  \bibinfo{author}{M.~C. Ridgway}, \bibinfo{author}{A.~P. Byrne},
  \bibinfo{author}{C.~Trautmann}, \bibinfo{author}{D.~J. Cookson},
  \bibinfo{author}{K.~Nordlund}, \bibinfo{author}{M.~Toulemonde},
\newblock \bibinfo{journal}{Phys. Rev. Lett.} \bibinfo{volume}{101}
  (\bibinfo{year}{2008}) \bibinfo{pages}{175503+}.
\bibitem[{Moreira et~al.(2010)Moreira, Devanathan, and
  Weber}]{Moreira2010Atomistic}
\bibinfo{author}{P.~A. F.~P. Moreira}, \bibinfo{author}{R.~Devanathan},
  \bibinfo{author}{W.~J. Weber},
\newblock \bibinfo{journal}{J. Phys. Condens. Matter} \bibinfo{volume}{22}
  (\bibinfo{year}{2010}) \bibinfo{pages}{395008+}.
\bibitem[{Wolf(1992)}]{Wolf1992Reconstruction}
\bibinfo{author}{D.~Wolf},
\newblock \bibinfo{journal}{Phys. Rev. Lett.} \bibinfo{volume}{68}
  (\bibinfo{year}{1992}) \bibinfo{pages}{3315--3318}.
\bibitem[{Wolf et~al.(1999)Wolf, Keblinski, Phillpot, and
  Eggebrecht}]{Wolf1999Exact}
\bibinfo{author}{D.~Wolf}, \bibinfo{author}{P.~Keblinski},
  \bibinfo{author}{S.~R. Phillpot}, \bibinfo{author}{J.~Eggebrecht},
\newblock \bibinfo{journal}{The Journal of Chemical Physics}
  \bibinfo{volume}{110} (\bibinfo{year}{1999}) \bibinfo{pages}{8254--8282}.
\bibitem[{Fennell and Gezelter(2006)}]{Fennell2006Ewald}
\bibinfo{author}{C.~J. Fennell}, \bibinfo{author}{J.~D. Gezelter},
\newblock \bibinfo{journal}{The Journal of Chemical Physics}
  \bibinfo{volume}{124} (\bibinfo{year}{2006}) \bibinfo{pages}{234104}.
\bibitem[{Zahn et~al.(2002)Zahn, Schilling, and Kast}]{Zahn2002Enhancement}
\bibinfo{author}{D.~Zahn}, \bibinfo{author}{B.~Schilling},
  \bibinfo{author}{S.~M. Kast},
\newblock \bibinfo{journal}{The Journal of Physical Chemistry B}
  \bibinfo{volume}{106} (\bibinfo{year}{2002}) \bibinfo{pages}{10725--10732}.
\bibitem[{Berendsen et~al.(1984)Berendsen, Postma, van Gunsteren, DiNola, and
  Haak}]{Berendsen1984Molecular}
\bibinfo{author}{H.~J.~C. Berendsen}, \bibinfo{author}{J.~P.~M. Postma},
  \bibinfo{author}{W.~F. van Gunsteren}, \bibinfo{author}{A.~DiNola},
  \bibinfo{author}{J.~R. Haak},
\newblock \bibinfo{journal}{The Journal of Chemical Physics}
  \bibinfo{volume}{81} (\bibinfo{year}{1984}) \bibinfo{pages}{3684--3690}.
\bibitem[{Lindan and Gillan(1991)}]{Lindan1991Molecular}
\bibinfo{author}{P.~J.~D. Lindan}, \bibinfo{author}{M.~J. Gillan},
\newblock \bibinfo{journal}{J. Phys.: Condens. Matter} \bibinfo{volume}{3}
  (\bibinfo{year}{1991}) \bibinfo{pages}{3929--3939}.
\bibitem[{Andersson and Backstrom(1987)}]{Andersson1987Thermal}
\bibinfo{author}{S.~Andersson}, \bibinfo{author}{G.~Backstrom},
\newblock \bibinfo{journal}{J. Phys. C: Solid State Phys.} \bibinfo{volume}{20}
  (\bibinfo{year}{1987}) \bibinfo{pages}{5951--5952}.
\bibitem[{Toulemonde et~al.(2000)Toulemonde, Dufour, Meftah, and
  Paumier}]{Toulemonde2000Transient}
\bibinfo{author}{M.~Toulemonde}, \bibinfo{author}{C.~Dufour},
  \bibinfo{author}{A.~Meftah}, \bibinfo{author}{E.~Paumier},
\newblock \bibinfo{journal}{Nucl. Instrum. Methods Phys. Res., Sect. B}
  \bibinfo{volume}{166-167} (\bibinfo{year}{2000}) \bibinfo{pages}{903--912}.
\bibitem[{Chen and Mart\'{\i}nez(2007)}]{Chen2007QTPIE}
\bibinfo{author}{J.~Chen}, \bibinfo{author}{T.~J. Mart\'{\i}nez},
\newblock \bibinfo{journal}{Chem. Phys. Lett.} \bibinfo{volume}{438}
  (\bibinfo{year}{2007}) \bibinfo{pages}{315--320}.
\bibitem[{Chadderton(2003)}]{Chadderton2003Nuclear}
\bibinfo{author}{L.~Chadderton},
\newblock \bibinfo{journal}{Radiat. Meas.} \bibinfo{volume}{36}
  (\bibinfo{year}{2003}) \bibinfo{pages}{13--34}.
\bibitem[{Jensen(1998)}]{Jensen1998Microscopic}
\bibinfo{author}{J.~Jensen},
\newblock \bibinfo{journal}{Nucl. Instrum. Methods Phys. Res., Sect. B}
  \bibinfo{volume}{146} (\bibinfo{year}{1998}) \bibinfo{pages}{399--404}.
\bibitem[{Li(2003)}]{Li2003AtomEye}
\bibinfo{author}{J.~Li},
\newblock \bibinfo{journal}{Modelling and Simulation in Materials Science and
  Engineering} \bibinfo{volume}{11} (\bibinfo{year}{2003})
  \bibinfo{pages}{173--177}.

\end{thebibliography}


\pagebreak

\pagebreak

\pagebreak

\pagebreak

\pagebreak

\end{document}